\documentclass[11pt,a4paper]{article}

\pdfoutput=1

\usepackage{graphicx}
\usepackage{comment}
\usepackage{epstopdf}
\usepackage{jcappub}
\usepackage{amsmath,amsthm,latexsym,amssymb,amsfonts}
\usepackage{fontenc,dsfont,layout}
\usepackage{hyperref}
\usepackage{psfrag}
\usepackage[cp1250]{inputenc}
\newcommand{\ee}{\end{equation}}
\newcommand{\br}{\begin{eqnarray}}
\newcommand{\bea}{\begin{eqnarray}}
\newcommand{\eea}{\end{eqnarray}}
\newcommand{\er}{\end{eqnarray}}
\newcommand{\ba}{\begin{array}}
\newcommand{\ea}{\end{array}}

\newcommand{\bi}{\begin{itemize}}
\newcommand{\ei}{\end{itemize}}
\newcommand{\bn}{\begin{enumerate}}
\newcommand{\en}{\end{enumerate}}
\newcommand{\bc}{\begin{center}}
\newcommand{\ec}{\end{center}}

\newcommand{\PRexp}{2.14 \pm 0.05}

\newcommand{\beq}{\begin{equation}}
\newcommand{\eeq}{\end{equation}}

\newcommand{\gsim}{\lower.7ex\hbox{$\;\stackrel{\textstyle>}{\sim}\;$}}
\newcommand{\lsim}{\lower.7ex\hbox{$\;\stackrel{\textstyle<}{\sim}\;$}}

\title{\centering Coleman-Weinberg linear inflation:\\ metric vs. Palatini formulation}

\author[a]{Antonio Racioppi}

\affiliation[a]{National Institute of Chemical Physics and Biophysics, R\"avala 10, 10143 Tallinn, Estonia}

\emailAdd{Antonio.Racioppi@kbfi.ee}

\abstract{It has been previously shown that the linear inflation appears naturally as a solution of Coleman-Weinberg inflation, provided that the inflaton has a non-minimal coupling to gravity and the Planck scale is dynamically generated. We revisit the previous study by improving the discussion of reheating and by comparing the results of the metric and the Palatini formulations of non-minimal gravity. We find that both formulations predict linear inflation but a different number of $e$-folds. If the non-minimal coupling is larger than one, future experimental sensitivity can discriminate between the two realizations.}


\begin{document}
\maketitle

\section{Introduction} \label{sec:Introduction}
Cosmic inflation~\cite{Starobinsky:1980te,Guth:1980zm,Linde:1981mu,Albrecht:1982wi} is a well established theory of the early universe.  Inflation addresses at the same time the fine-tuning problems of the hot big bang (namely the horizon and flatness problems) and produces a power spectrum of primordial inhomogeneities of cosmic microwave background (CMB) in good agreement with 
several experiments \cite{Ade:2015tva,Ade:2015xua,Ade:2015lrj,Array:2015xqh}. 
The latest data from the BICEP2/Keck collaboration \cite{Array:2015xqh} strongly constrains the tensor-to-scalar ratio $r$, which tells us about the amplitude of primordial gravitational waves and about the scale of inflation. In particular, the predictions of linear inflation for $r$ in function of the scalar spectral index ($n_s$) lie on the very edge of the $2 \sigma$ limit of the data best fit, leaving linear inflation as the first model to be confirmed or ruled out by the next data release.

In \cite{Rinaldi:2015yoa,Kannike:2015kda,Barrie:2016rnv,Artymowski:2016dlz} it has been shown that the linear inflaton potential appears naturally\footnote{An alternative mechanism relying on fermion condensates is presented in Ref.~\cite{Iso:2014gka}.}  as a solution of Coleman-Weinberg~\cite{Coleman:1973jx} (CW) inflation, provided that the inflaton has a non-minimal coupling to gravity and the Planck scale is dynamically generated.
The concept of CW inflation was already introduced in the early papers \cite{Linde:1981mu,Albrecht:1982wi,Linde:1982zj,Ellis:1982ws,Ellis:1982dg}.
Recently this class of models became again popular \cite{Kannike:2014mia,Kannike:2015apa,Marzola:2015xbh,Marzola:2016xgb,Kannike:2016wuy,
Rinaldi:2015uvu,Farzinnia:2015fka,Karananas:2016kyt,Tambalo:2016eqr,Kaneta:2017lnj,Saadi:2017dns}, pushed by the Higgs boson discovery \cite{Aad:2012tfa,Chatrchyan:2012xdj} and by the subsequent quest (of a relevant part of the scientific community) for new symmetries to explain naturalness in absence of supersymmetry, like classical scale invariance \cite{Bardeen:1995kv,Heikinheimo:2013fta}. Such class of theories is based on dimensional transmutation \cite{Coleman:1973jx} and dynamical generation of mass scales via CW potentials induced by new scalar particles \cite{Hempfling:1996ht,Gabrielli:2013hma,Marzola:2015xbh,Marzola:2016xgb,Marzola:2017jzl}. Therefore, the application of such theories to inflation is quite straightforward. In this paper we study this type of inflation models in the presence of a non-minimal coupling to gravity \cite{Bezrukov:2007ep,Bezrukov:2010jz}.

A modification of general relativity (GR) requires a discussion of what are the gravitational degrees of freedom. In the usual metric formulation of GR \cite{Padmanabhan:2004fq,Paranjape:2006ca} the independent variables are the metric and its first derivatives, while in the Palatini formulation \cite{Palatini:1919,Einstein:1925,Ferraris:1982} the independent variables are the metric and the connection. Given the Einstein-Hilbert Lagrangian, the two formalism share the same equations of motion and therefore describe equivalent physical theories \cite{Exirifard:2007da} (up to a boundary term in the action \cite{PhysRevLett.28.1082,PhysRevD.15.2752}). However, if  non-minimal couplings between gravity and matter are introduced, such equivalence is lost and metric and Palatini formulations describe two different theories of gravity \cite{Bauer:2008zj}. Recently several authors raised again interest in this topic \cite{Bauer:2010bu,Tamanini:2010uq,Bauer:2010jg,Olmo:2011uz,Markkanen:2014dba,
Rasanen:2017ivk,Tenkanen:2017jih}.

The aim of this work is to extend the previous discussion of the non-minimal CW model of \cite{Kannike:2015kda}, presenting a comparative analysis of the possible gravity formulation (metric or Palatini) and a more precise study of reheating. 

The article is organized as follows.
In section~\ref{sec:Preliminaries} we set the notation reintroducing the main concepts about the effective potential, the Palatini formulation of a gravity theory and their application to our specific model. In section \ref{sec:inflation} we present the comparative study of inflationary predictions in both metric and Palatini formulation, including the reheating phenomenology.
We conclude in section~\ref{sec:Summary}. Technical details are then presented in Appendices~\ref{appendix:potential} and \ref{appendix:linear:limit}.

\section{Model building} \label{sec:Preliminaries}
Our interest is to study classical (quasi-)scale invariant theories, therefore the general form for the Lagrangian in the Jordan frame is:
\begin{equation}
  \sqrt{- g^{J}} \mathcal{L}^{J} = \sqrt{- g^{J}} \left[ \mathcal{L}_{\rm SM} - \frac{\xi_\phi}{2} \phi^{2} R \left( \Gamma \right)
  + \frac{(\partial \phi)^{2}}{2} - V^J_{\rm 1-loop}(\phi) + \mathcal{L}^{J}(\sigma, \psi, A_\mu) + \Lambda^4\right] ,
   \label{eq:Jordan:Lagrangian}
\end{equation}
where $\mathcal{L}_{\rm SM} $ is the SM Lagrangian, $R$ is the Ricci scalar constructed from a connection $\Gamma$, $V^J_{\rm 1-loop}(\phi)$ is a generic 1-loop scalar potential\footnote{{ It has been demonstrated that the cosmological perturbations are invariant under
a change of frame (see for instance \cite{Prokopec:2013zya,Jarv:2016sow} and refs. therein). On the other side, the quantum level equivalence of the Einstein and Jordan
frames is still an open issue. In the present article we adopt the following computational strategy: we compute (or assume) that the effective potential in the Jordan frame is given by eq. (\ref{eq:Veff:Jordan:Lambda}). Once we got the full expression of the 1-loop Jordan frame scalar potential, we move to the Einstein frame, where the computation of the slow-roll parameters is easier. Given a scalar potential in the Jordan frame, the cosmological perturbations are then independent, in the slow-roll approximation, from the choice of the frame in which we perform the inflationary computation \cite{Prokopec:2013zya,Jarv:2016sow}.} For further readings on frame equivalence and/or loop corrections in scalar-tensor theories see Refs. \cite{Jarv:2014hma,Kuusk:2015dda,Kuusk:2016rso,Flanagan:2004bz,Catena:2006bd,Barvinsky:2008ia,DeSimone:2008ei,
Barvinsky:2009fy,Steinwachs:2011zs,Chiba:2013mha,George:2013iia,Postma:2014vaa,
Kamenshchik:2014waa,George:2015nza,Miao:2015oba,Burns:2016ric,Hamada:2016onh,Karam:2017zno}.} generated from a classical scale invariant setup,\footnote{In presence of an Einstein-Hilbert term $-\frac{M^2}{2} R$, the Lagrangian (\ref{eq:Jordan:Lagrangian}) can be considered as a limiting case $M \ll M_P$, which can be easily realised with the \emph{natural} choice $M \sim \Lambda$ (cf. Fig. \ref{Fig:Results:before}c).}
 \begin{equation}
  V^J_{\rm 1-loop}(\phi) = \frac{1}{4} \lambda_\phi (\phi) \phi^4 \label{eq:Veff:J},
 \end{equation}
 and $\mathcal{L}^{J}(\sigma, \psi, A_\mu)$ are the Lagrangian terms involving the extra particle content that is generating the dominant contribution to the inflaton effective potential $V^J_{\rm 1-loop}(\phi)$. For the purpose of this article we do not need to specify the content of $\mathcal{L}^{J}(\sigma, \psi, A_\mu)$ (we will briefly comment about it later), but we simply assume that it satisfies the classical scale invariance requirement.
 The Lagrangian (\ref{eq:Jordan:Lagrangian}) lacks of an Einstein-Hilbert term. This has to be generated by inducing a non vanishing vacuum expectation value (VEV) of the inflaton. Therefore, to generate the Planck scale, the VEV $v_\phi$ of the inflaton field must be given by
\begin{equation}
  v_{\phi}^{2} = \frac{M_P^{2}}{\xi_\phi}.
  \label{eq:v:phi:Planck:mass}
\end{equation}
Note that such a relation automatically implies that $\xi_\phi$ can only take positive values. 
The cosmological constant $\Lambda$ is adjusted so that at the minimum the potential value is zero, i.e.,
 \begin{equation}
  V_\text{eff}(v_\phi)=\frac{1}{4} \lambda_\phi (v_\phi) v_\phi^4 + \Lambda = 0 \label{eq:Vmin} \, .
 \end{equation}
We will see later that inflationary constraints will impose $\Lambda \ll M_P$ (see section \ref{sec:inflation}) and therefore classical scale invariance will be only softly broken.


Let us discuss now in more details the quartic coupling $\lambda$. Our focus is on the simplest\footnote{The non-minimal coupling is generally subject to loop corrections parametrized by a beta-function of the following type
\begin{equation}
  16 \pi^2 \beta_{\xi_\phi} \approx \xi_\phi \sum_k \lambda_k  \, ,
  \label{eq:betaxi}
\end{equation}
where $\sum_k \lambda_k$ represents some other couplings from the scalar potential, for example the ones contained in $\mathcal{L}^{J}(\sigma, \psi, A_\mu)$ that are also generating the running of $\lambda_\phi$. In order to ignore the quantum corrections to $\xi_\phi$, the condition $\beta_{\xi_\phi} \ll \xi_\phi $ must be satisfied. This has been explicitly realized in \cite{Kannike:2015apa,Kannike:2015kda,Marzola:2015xbh,Marzola:2016xgb}.
However, because of the constraint on the amplitude of scalar perturbations (\ref{eq:As}),
perturbativity of the theory and the $16 \pi^2$ suppression factor, we assume that such condition holds also for our model independent construction.} CW configuration, i.e., we assume that 
\begin{equation}
 \lambda_\phi (\phi) \simeq \lambda_\phi(v_\phi) + \beta_{\lambda _{\phi}}(v_\phi) \ln\frac{\phi }{v_\phi} \, , \label{eq:lambda:run}
\end{equation}
is a good approximation of the quartic coupling in eq. (\ref{eq:Veff:J}), where $\beta_{\lambda_\phi} (\mu)= \mu \frac{\partial}{\partial \mu}\lambda_\phi (\mu)$ is the beta-function\footnote{We are making the implicit assumption that the contribution of $\lambda_\phi$ to its running is subdominant and that the main contribution is coming from $ \mathcal{L}^{J}(\sigma, \psi, A_\mu)$. The minimal explicit model involves an extra scalar \cite{Kannike:2015kda} that can play the role of super-heavy dark matter \cite{Kannike:2016jfs}. However, for the present model independent discussion, in order to keep the approximation in eq. (\ref{eq:lambda:run}) valid, we just need to assume that the dominant contribution is coming from bosonic degrees of freedom. For more details about the possible configurations check Appendix \ref{appendix:potential}.}  of the quartic coupling $\lambda_\phi (\mu)$. By using eqs. (\ref{eq:v:phi:Planck:mass}), (\ref{eq:Vmin}) and (\ref{eq:lambda:run}) and imposing that $v_\phi \neq0 $ is the inflation VEV, it is possible to show \cite{Kannike:2015kda} that the potential can be rewritten as
\begin{eqnarray}
  V_{\rm eff}(\phi)&=& \frac{1}{4} \lambda_\phi (\phi) \phi^4 + \Lambda^4 = \Lambda ^4 \left\{ 1 + \left[ 2 \ln \left(\frac{\phi^2}{v_\phi^2}\right) -1 \right] \frac{\phi^4}{v_\phi^4} \right\} \label{eq:V:CW}\\
&=& \Lambda ^4 \left\{ 1 + \left[ 2 \ln \left(\frac{\xi_\phi \phi^2}{M_P^2}\right) -1 \right] \frac{\xi_\phi^2 \phi^4}{M_P^4} \right\} \label{eq:Veff:Jordan:Lambda}.
\end{eqnarray}

The inflationary predictions of the CW potential (\ref{eq:V:CW}), in absence of a non-minimal coupling to gravity, are well known and compatible with data only at a 2-$\sigma$ level in a reduced region of the parameters space \cite{Ade:2015tva,Ade:2015lrj,Array:2015xqh}. Such predictions are dramatically changed if a non-minimal coupling to gravity is added, as we do in Lagrangian (\ref{eq:Jordan:Lagrangian}). However, a modification of gravity calls for a discussion of what theory of gravity we are going to consider. This will be shortly discussed in the following subsection.

\subsection{Non-minimal Palatini gravity} \label{sec:Non_min_PG}
In this section, we give a brief discussion of the Palatini formulation of gravity\footnote{To the reader interested in further details on the topic, we suggest \cite{Bauer:2008zj} and references therein.} stressing, when needed, the differences with the corresponding metric formulation.  
The properties of space-time are essentially described by two quantities: a connection and a metric tensor.
The connection, ${\Gamma}^{\lambda}_{\alpha\beta}$, describes the parallel transport of tensor fields along a given curve.  If the space-time is curved, parallel transport around a closed path, after a full cycle, results in a finite mismatch. The curvature is uniquely determined by the Riemann tensor $R^{\mu}_{\alpha \nu \beta}\left(\Gamma\right)$ whose contraction $R_{\alpha \beta}\left(\Gamma\right) \equiv R^{\mu}_{\alpha \mu \beta}\left(\Gamma\right)$ gives the Ricci tensor\footnote{The connection determines not only the curvature but also the twirl of the space-time. The latter is encoded in the torsion tensor \cite{Hehl:1994ue}. However in this article we are going to study only torsion-free configurations. This is achieved by considering only symmetric connections $\Gamma^{\lambda}_{\alpha \beta} = \Gamma^{\lambda}_{\beta \alpha}$.}. Then, the metric tensor, $g_{\mu \nu}$, allows us to introduce the notion of distance and angles. The connection coefficients and metric tensor are fundamentally independent quantities. They exhibit {\it a priori} no known relationship, and if they are to have any it must be derived from an additional constraint or geometrodynamics \cite{Palatini:1919,Einstein:1925,Ferraris:1982}. The metric formulation consists in replacing $\Gamma$ with the Levi-Civita connection $\bar\Gamma$
\begin{eqnarray}
\label{eq:LC}
\overline{\Gamma}^{\lambda}_{\alpha \beta} = \frac{1}{2} g^{\lambda \rho} \left( \partial_{\alpha} g_{\beta \rho}
+ \partial_{\beta} g_{\rho \alpha} - \partial_{\rho} g_{\alpha \beta}\right) \, ,
\end{eqnarray}
while the Palatini formulation treats the connection as an independent variable whose behaviour is set by its equation of motion. In GR the two formalisms are equivalent. 

Let us consider now the non-minimally coupled theory described in eq. (\ref{eq:Jordan:Lagrangian}). Under the metric formalism the connection is the Levi-Civita one computed with the Jordan frame metric. However, using the Palatini formalism, the connection will have a different expression. Solving the equation of motion, we get \cite{Bauer:2008zj}
\begin{eqnarray}
\Gamma^{\lambda}_{\alpha \beta} = \overline{\Gamma}^{\lambda}_{\alpha \beta}
+ \delta^{\lambda}_{\alpha} \partial_{\beta} \omega(\phi) +
\delta^{\lambda}_{\beta} \partial_{\alpha} \omega(\phi) - g_{\alpha \beta} \partial^{\lambda}  \omega(\phi) \, ,
\label{eq:conn:J}
\end{eqnarray}
where
\begin{eqnarray}
\label{omega}
\omega\left(\phi\right)=\ln\sqrt{\frac{\xi_\phi \phi^2}{M_P^2}} \, .
\end{eqnarray}
Being the connections (\ref{eq:LC}) and (\ref{eq:conn:J}) different, the metric and Palatini formulation provide indeed two different theories of gravity.
Another way of seeing the differences is to change the frame by means of 
\begin{eqnarray}
\label{eq:gE}
g_{\mu \nu}^E = \Omega(\phi)^2 \ g_{\mu \nu} \, , \qquad \Omega(\phi)^2=e^{2 \omega(\phi)} \, ,
\end{eqnarray}
and study the problem in the Einstein frame. In the Einstein frame gravity looks the same in the two formalisms (see also eq. (\ref{eq:conn:J})), however the matter sector (in particular $\phi$) behaves differently. Performing the computations \cite{Bauer:2008zj}, the Einstein frame Lagrangian becomes
\begin{equation}
  \sqrt{- g^{E}} \mathcal{L}^{E} = 
  \sqrt{- g^{E}} \Bigg[  - \frac{M_P^2}{2} R  + 
  \frac{(\partial \phi_E)^{2}}{2} - \frac{V^J_{\rm eff}}{\Omega^4} + \dots \Bigg] \, ,
   \label{eq:Einstein:Lagrangian}
\end{equation}
where the ellipsis stands for the SM and extra matter Lagrangian transformed in the Einstein frame and the canonically normalized scalar field is\footnote{This result is not general but is a consequence of the choice of the non-minimal coupling in eq. (\ref{eq:Jordan:Lagrangian}). }
\begin{equation}
\phi_E = \frac{M_P}{q} \ln \frac{\phi}{v_\phi} \, ,
  \label{eq:dphiE}
\end{equation}
where $q$ is either
\begin{equation}
q = q_m = \sqrt{\frac{\xi_\phi}{1+ 6 \xi_\phi}} \, ,
  \label{eq:q:metric}
\end{equation}
in case of the metric formulation, or 
\begin{equation}
q = q_P = \sqrt{\xi_\phi} \, ,
  \label{eq:q:Palatini}
\end{equation}
in case of the Palatini formulation. 
Therefore we can see that the difference between the two formulations in the Einstein frame relies in the different non-minimal kinetic term involving $\phi$, or after the corresponding field redefinition, in different $\phi_E$ interactions.

\section{Inflationary results} \label{sec:inflation}
In this section we compute the inflationary predictions in the Einstein frame for the two different gravity formulations. First of all we will present model-independent results considering the usual range for the number of $e$-folds ($50 \leq N_e \leq 60$). Then we will improve the results by considering the most minimal reheating scenario with the inflaton decaying to SM particles via the non-minimal coupling to gravity.
\subsection{General computations}
\begin{figure}[t!]
\begin{center}
 \includegraphics[width=0.45\textwidth]{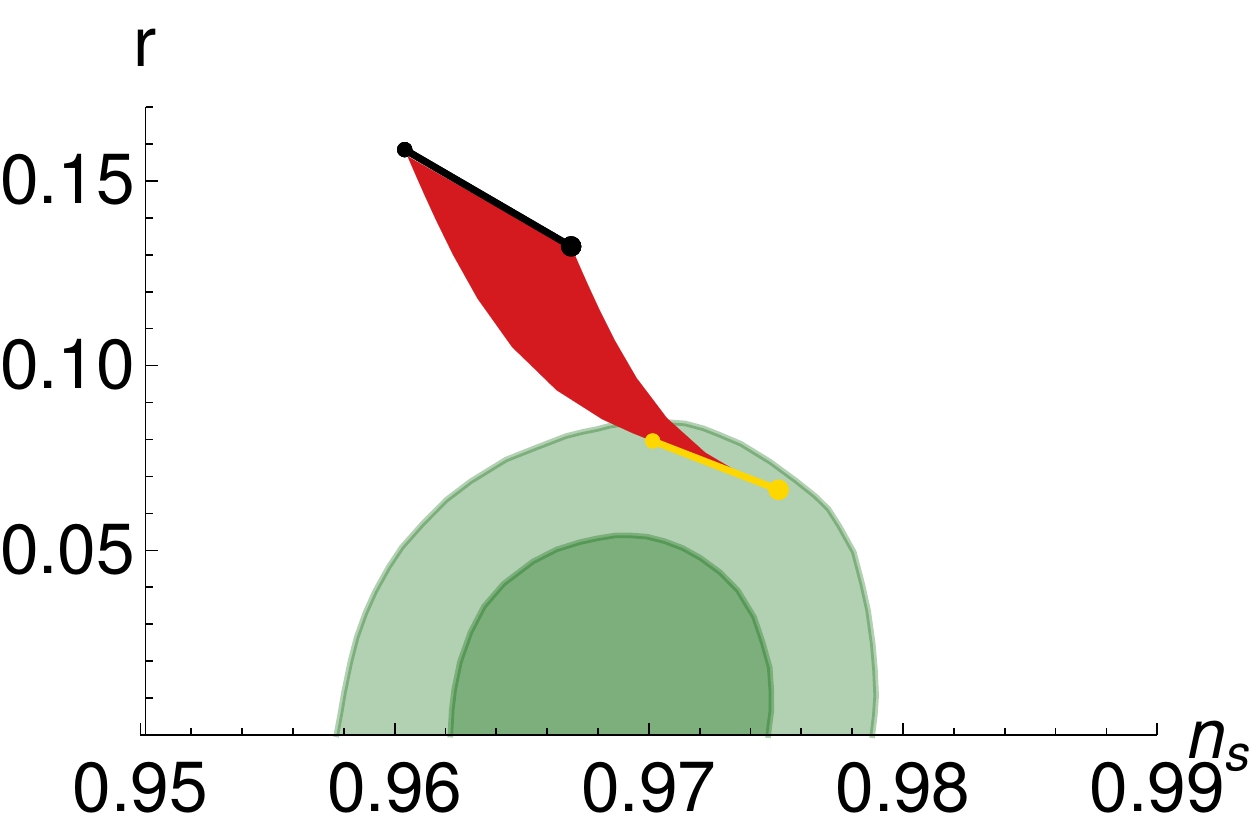}
 \includegraphics[width=0.45\textwidth]{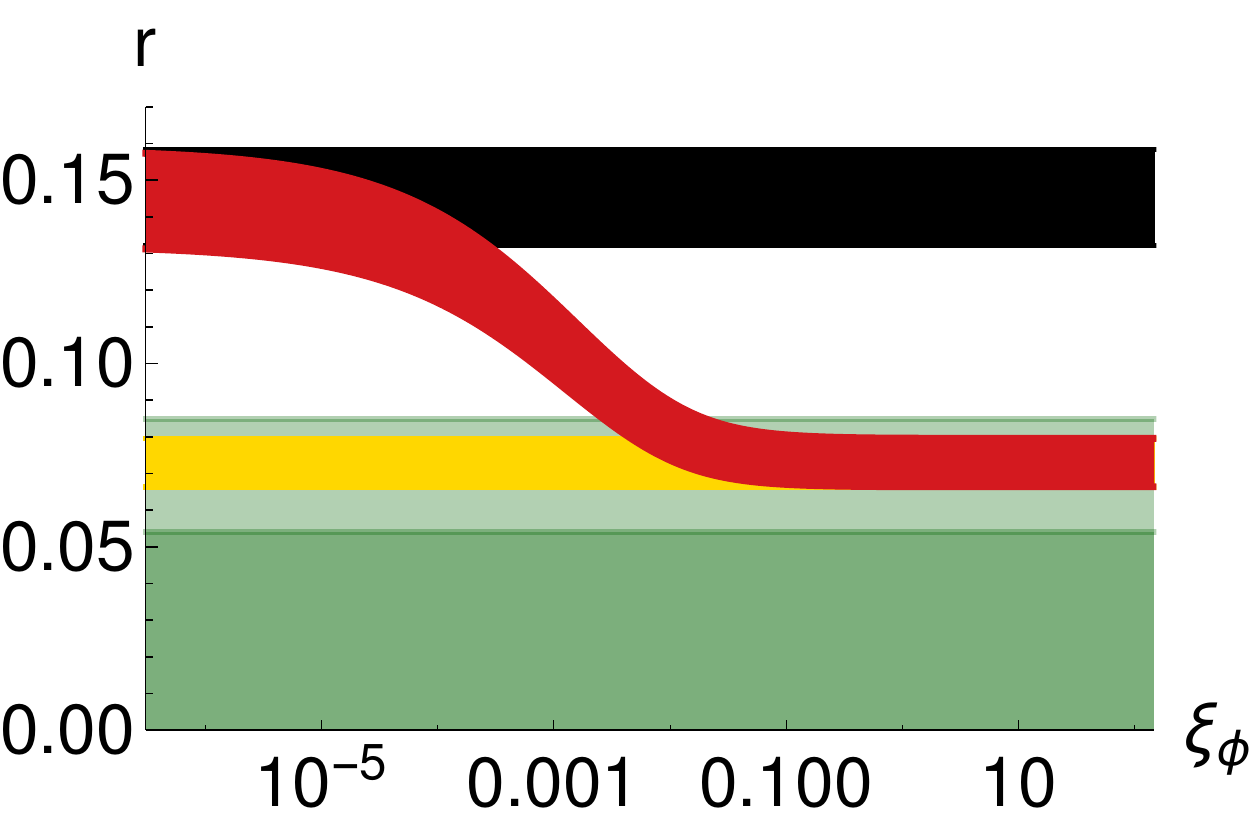}\\
 a) \hspace{0.45\textwidth} b) \\
 \includegraphics[width=0.45\textwidth]{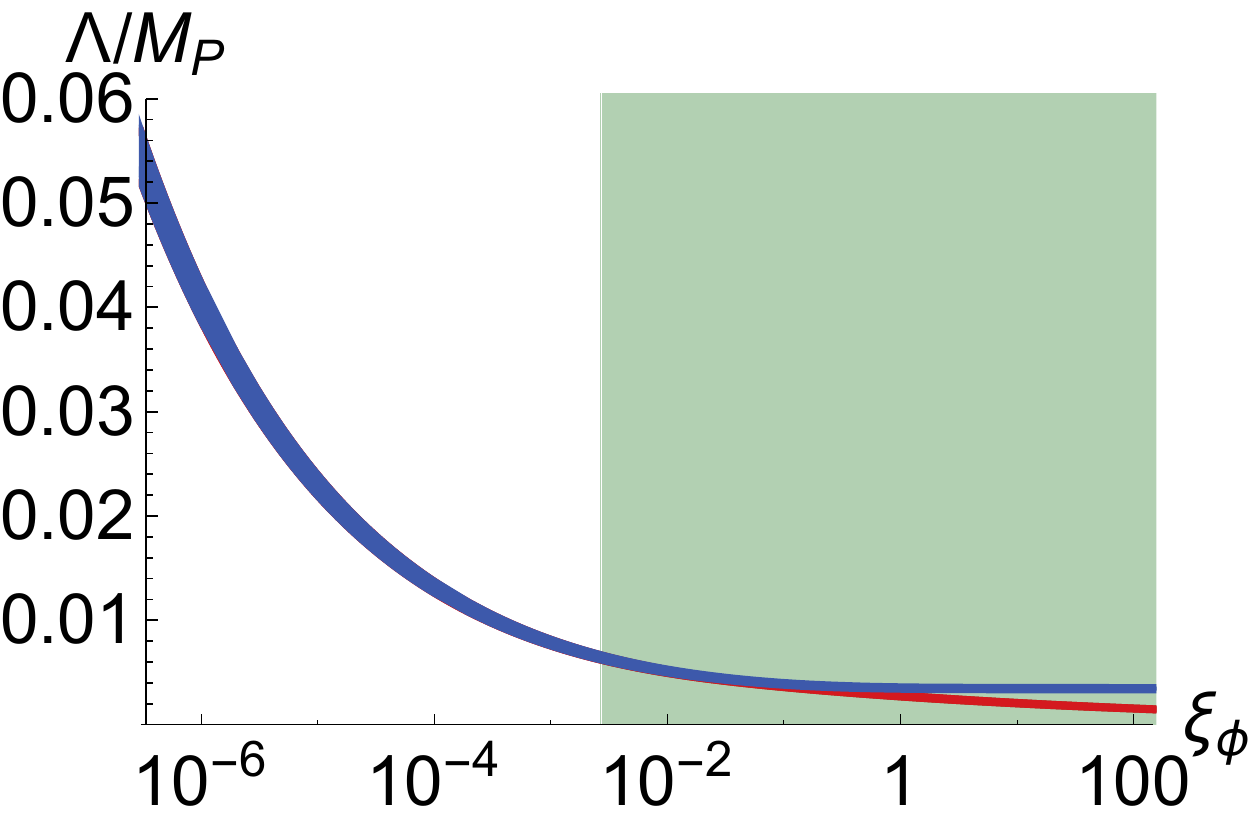}
 \includegraphics[width=0.45\textwidth]{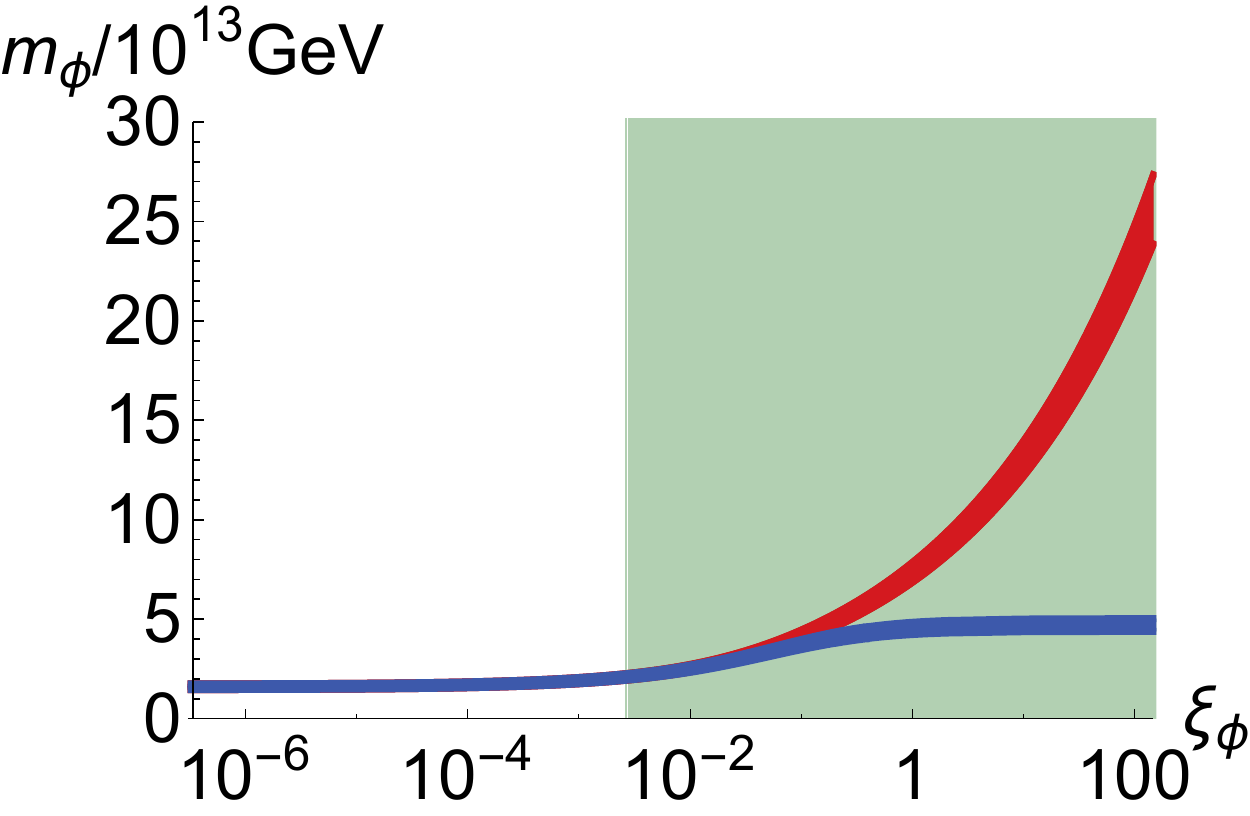}\\
  c) \hspace{0.45\textwidth} d)
\end{center}
 \caption{Tensor-to-scalar ratio $r$ for the metric (blue) and the Palatini (red) formulation with $N_e \in [50,60]$ $e$-folds as a function of $n_s$ (a) and, as a function of $\xi_\phi$, $r$ (b), $\Lambda$ (c) and $m_\phi$ (d). For reference we also plot predictions of  $m^2 \phi_E^2$ (black) and $m^3 \phi_E$ (yellow) potentials. The light green areas present the 1,2 $\sigma$ constraints from the BICEP2/Keck data \cite{Array:2015xqh}. This figure is adapted with permission from \cite{Array:2015xqh}.}
  \label{Fig:Results:before}
\end{figure}

Given the Einstein frame Lagrangian (\ref{eq:Einstein:Lagrangian}) and using eqs. (\ref{eq:Veff:Jordan:Lambda}), (\ref{eq:gE}) and (\ref{eq:dphiE}), the corresponding scalar potential is
\begin{equation}
V_E(\phi_E)= \frac{V^J_{\rm eff}}{\Omega^4}=\Lambda ^4 \left(4 \, q \frac{\phi_E}{M_P}+e^{-4 \, q \frac{\phi_E}{M_P}}-1\right) \, . \label{eq:Veff:Einstein}
\end{equation}
The potential has the same shape under both formulations. The only difference is the pre-factor $q$ in front of the $\frac{\phi_E}{M_P}$ terms.
Using the slow-roll approximation we have that the tensor-to-scalar ratio $r$ and the scalar spectral index $n_s$ are given by
\begin{eqnarray}
r &=& 16 \epsilon, \label{eq:r}\\
n_s &=& 1 -6 \epsilon + 2 \eta \label{eq:n_s},
\end{eqnarray}
where $\epsilon$ and $\eta$ are the slow-roll parameters defined as
\begin{eqnarray}
\epsilon &=& \frac{M_P^2}{2} \left( \frac{V_E'}{V_E} \right)^2,\\
\eta &=& \frac{M_P^2}{2} \frac{V_E''}{V_E},
\end{eqnarray}
where, according to the configuration of interest, $V_E$ is the inflaton potential given by eq. (\ref{eq:Veff:Einstein}), and the symbol `` $'$ '' denotes differentiation with respect to $\phi_E$. The corresponding predictions are given in Fig. \ref{Fig:Results:before}. We present the results\footnote{The discussion is pretty similar to the one in \cite{Kannike:2015kda}, therefore here we just mention the relevant features. The interested reader can find the detailed analysis in \cite{Kannike:2015kda}.} for the tensor-to-scalar ratio $r$ for the metric (blue) and the Palatini (red) formulation with $N_e \in [50,60]$ $e$-folds as a function of $n_s$ (\ref{Fig:Results:before}a) and as a function of $\xi_\phi$ (\ref{Fig:Results:before}b). For reference we also plot predictions of quadratic (black) and linear (yellow) potentials.
The light green areas present the 1,2 $\sigma$ constraints from the BICEP2/Keck data \cite{Array:2015xqh}.
The potential (\ref{eq:Veff:Einstein}) has formally only two parameters: $\Lambda$, that should be fixed by normalisation, and $\xi_\phi$ that is the only free dynamical parameter. Since the slow-roll parameters, in particular the observables $r$ and $n_s,$ are normalisation independent, the results of Fig.~\ref{Fig:Results:before} depend only on the value of $\xi_\phi$ (and on $N_e$).
We notice that in Figs.~\ref{Fig:Results:before}a and \ref{Fig:Results:before}b seems impossible to distinguish between the two cases\footnote{For more details see Appendix \ref{appendix:linear:limit}.}, being the red region over the blue one. 
This result is not general\footnote{For instance in the usual Higgs-inflation model, it is possible to clearly distinguish between the formulations \cite{Bauer:2008zj,Rasanen:2017ivk}.} but is a consequence of the particular choice of the non-minimal coupling in eq. (\ref{eq:Jordan:Lagrangian}). 
For very small values of $\xi_\phi$ the predictions essentially coincide with the ones of quadratic inflation. By increasing $\xi_\phi$, $r$ decreases, saturating the linear limit
for $\xi_\phi\gtrsim 0.1$.

To complete the model independent analyses, we discuss the constant $\Lambda$ and the inflaton mass $m_\phi$.
$\Lambda$ is fixed in order to agree with the constraint on the amplitude of scalar perturbations \cite{Ade:2015lrj,Ade:2015xua}
\begin{equation}
 A_s=(\PRexp)\times 10^{-9} \, . \label{eq:As}
\end{equation}
The results for $\Lambda$ as a function of $\xi_\phi$ are presented in Fig. \ref{Fig:Results:before}c. As expected, $\Lambda$ is always sub-Planckian. In the metric formulation, for $\xi_\phi$ increasing, $\Lambda$ approaches an asymptotic value $\Lambda_\infty=M_P \sqrt[4]{A_s}  \frac{3^{3/8}}{2 \left(4 N_e+1\right){}^{3/8}}$, while in the Palatini case is always decreasing (see also Appendix \ref{appendix:linear:limit}).
The inflaton mass can be computed from potential (\ref{eq:Veff:Einstein}) as
\begin{equation}
 m_\phi^2 = \left. V_E''\right|_{\phi_E=0} = 16 q^2 \frac{\Lambda ^4}{M_P^2} \, .
\end{equation}
We present  in Fig.~\ref{Fig:Results:before}d the inflaton mass $m_\phi$ as a function of $\xi_\phi$. In analogy to what happens to $\Lambda$,  in the metric formulation, for $\xi_\phi$ increasing, $m_\phi^2$ approaches an asymptotic value $m^2_\infty= A_s \, M_P^2 \frac{4 \sqrt{3} \pi }{\left(4 N_e+1\right){}^{3/2}}$, while in the Palatini case is always increasing.

To conclude we notice that, in practise, only the linear limit region lies inside the favoured region of BICEP2/Keck data, therefore our further studies will be focused on the linear limit case.

\subsection{Minimal model of reheating} \label{section:after:reheating}
As clear from Figs.~\ref{Fig:Results:before}a and \ref{Fig:Results:before}b, it seems impossible to distinguish between the metric and the Palatini formulation. However, it is possible to discriminate between the two formulations once reheating is taken into account. For simplicity, let us work in the Einstein frame. The inflaton can decay into SM particles via the non-minimal coupling to gravity, for example into Higgs boson pairs.  The full theory Lagrangian contains, of course, also the SM terms,\footnote{Electroweak symmetry breaking may be induced in a classically scale invariant theory as, for  instance, in \cite{Gabrielli:2013hma}. However we do not specify such a mechanism here since it is beyond the aim of this work.} plus eventual BSM physics. After moving from the Jordan frame to the Einstein frame, the Higgs kinetic term (ignoring now for simplicity the gauge term of the covariant derivative) will be
\begin{equation}
\begin{split}
 S_{\text{kin},h} &=\int d^4 x \sqrt{-g^E} \frac{\partial_\mu \left( h_E \Omega \right) \partial^\mu \left( h_E \Omega \right)}{2\Omega^2}
 \\
 &\simeq  \int d^4 x \sqrt{-g^E} \Big( \frac{1}{2} \partial_\mu h_E \partial^\mu h_E +
           q \frac{h_E \partial_\mu {\phi_E} \partial^\mu h_E}{M_{\rm P}}+ \dots \Big),
\end{split}
\label{eq:chihhvertex}
\end{equation}
where $q$, as usual, differs according to if we are using the metric or the Palatini formulation, we expanded $\Omega(\phi({\phi_E}))$ for ${\phi_E} \ll M_{\rm P}$ and kept only the leading order correction. The last term of eq. (\ref{eq:chihhvertex}) will induce the inflaton decay into a pair of Higgs scalars.\footnote{A contribution to it was computed in~\cite{Csaki:2014bua}.}
The decay width for the process is
\begin{equation}
 \Gamma_{\phi h h} = \frac{q^2}{64 \pi} \frac{m_\phi^3}{ M_{\rm P}^2},
\end{equation}
where we neglected the Higgs mass since $m_h \ll m_\phi$.
The kinetic term of the SM fermions and gauge vectors is invariant under conformal transformation, therefore from there we cannot get a similar contribution. However, Goldstone bosons have the same contribution. Therefore, we can induce an inflaton decay to vectors through a decay into the Goldstone bosons obtaining
\begin{equation}
 \Gamma_{\phi Z Z} = \frac{1}{2} \Gamma_{\phi W W} = \Gamma_{\phi h h}.
\end{equation}
The same procedure can be extended to all the scalar particles of the theory, included the ones in the hidden sector. However in our minimal scenario, we neglect the possibility of such decay channels, so that
\begin{equation}
 \Gamma_{\phi} = 4 \Gamma_{\phi h h} .
\end{equation}

\begin{figure}[t!]
\begin{center}
 \includegraphics[width=0.45\textwidth]{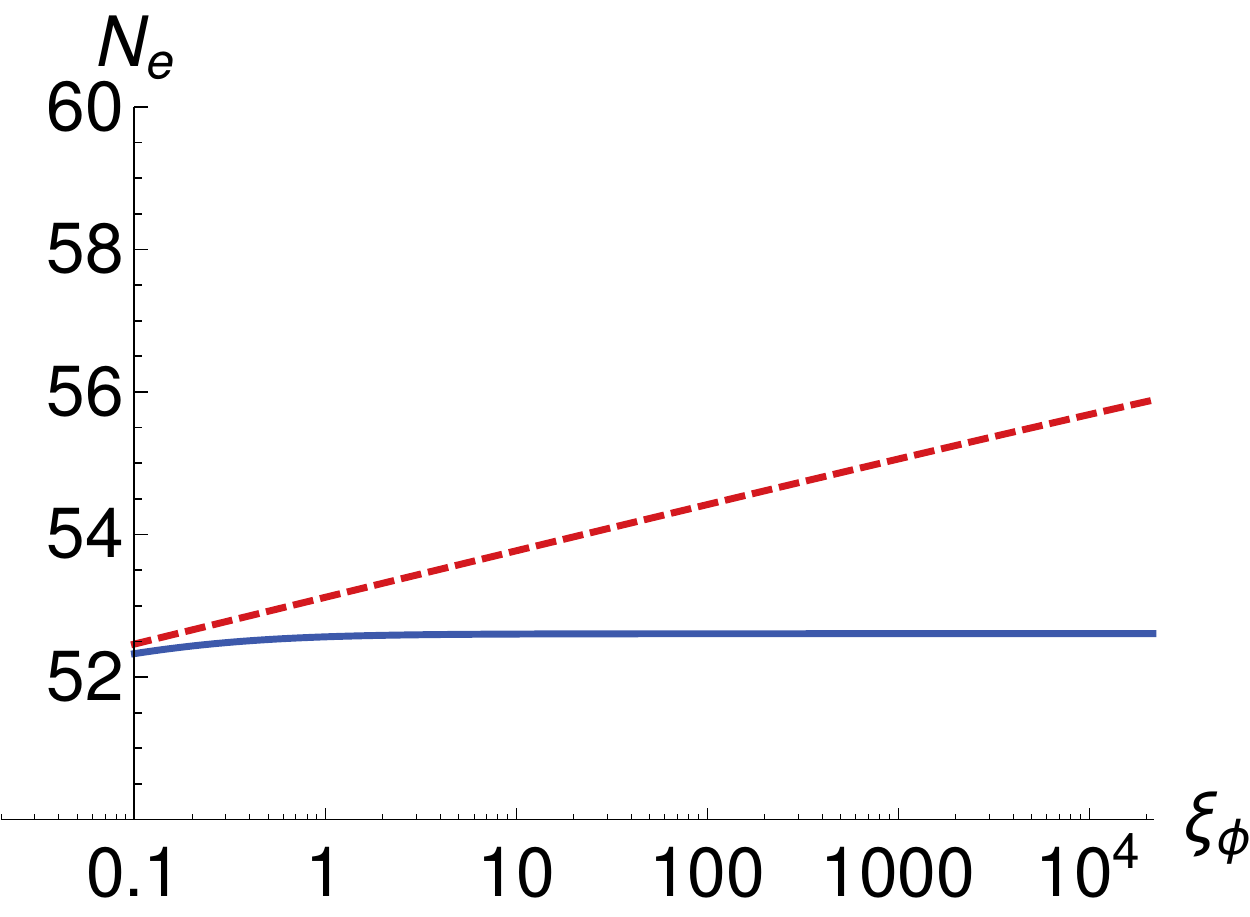} 
 \includegraphics[width=0.45\textwidth]{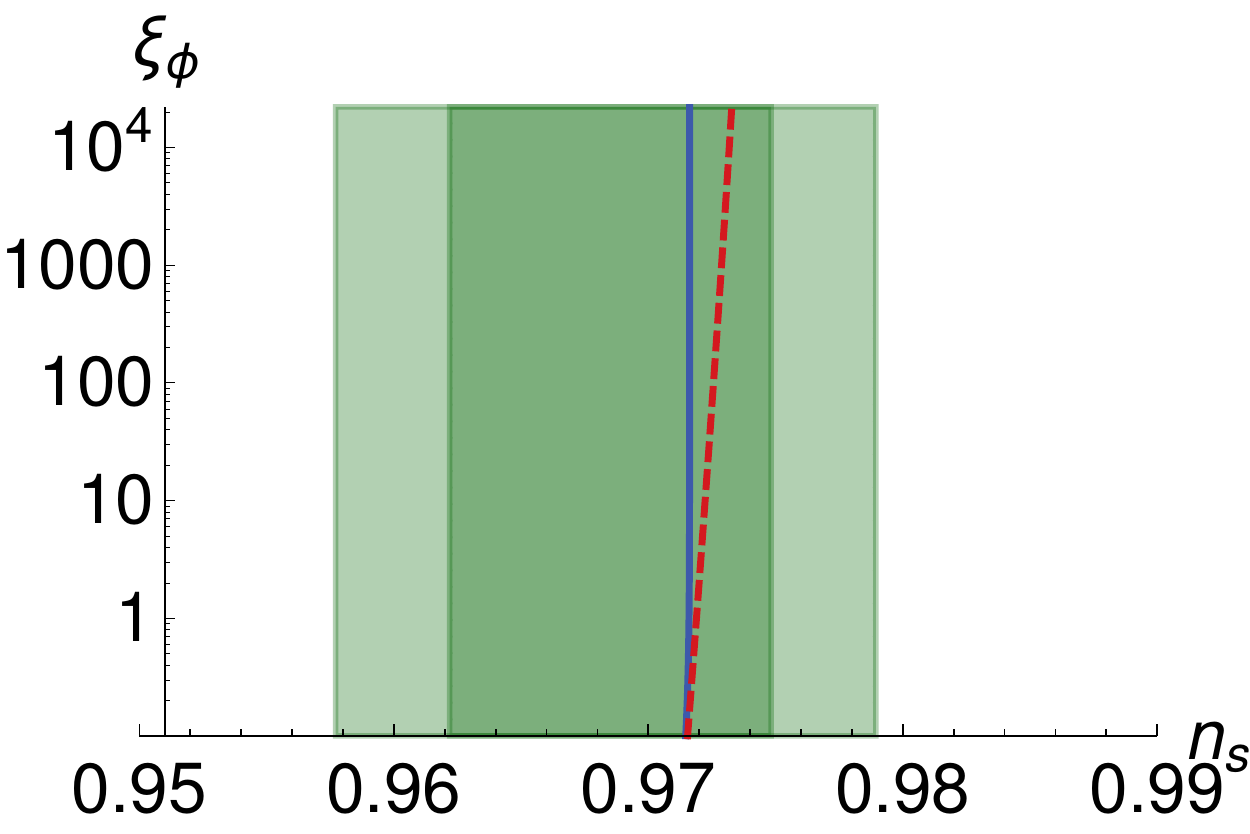}\\
 a) \hspace{0.45\textwidth} b) \\
 \includegraphics[width=0.45\textwidth]{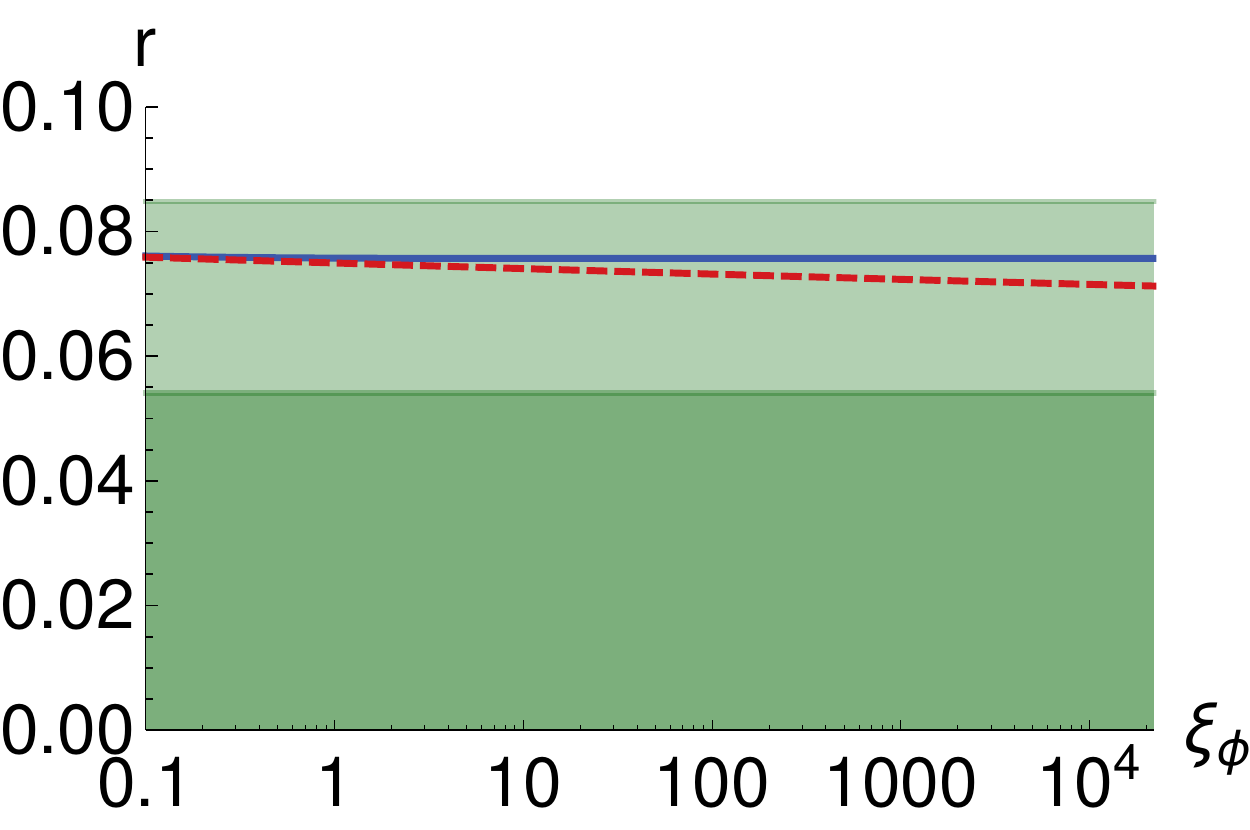}
 \includegraphics[width=0.45\textwidth]{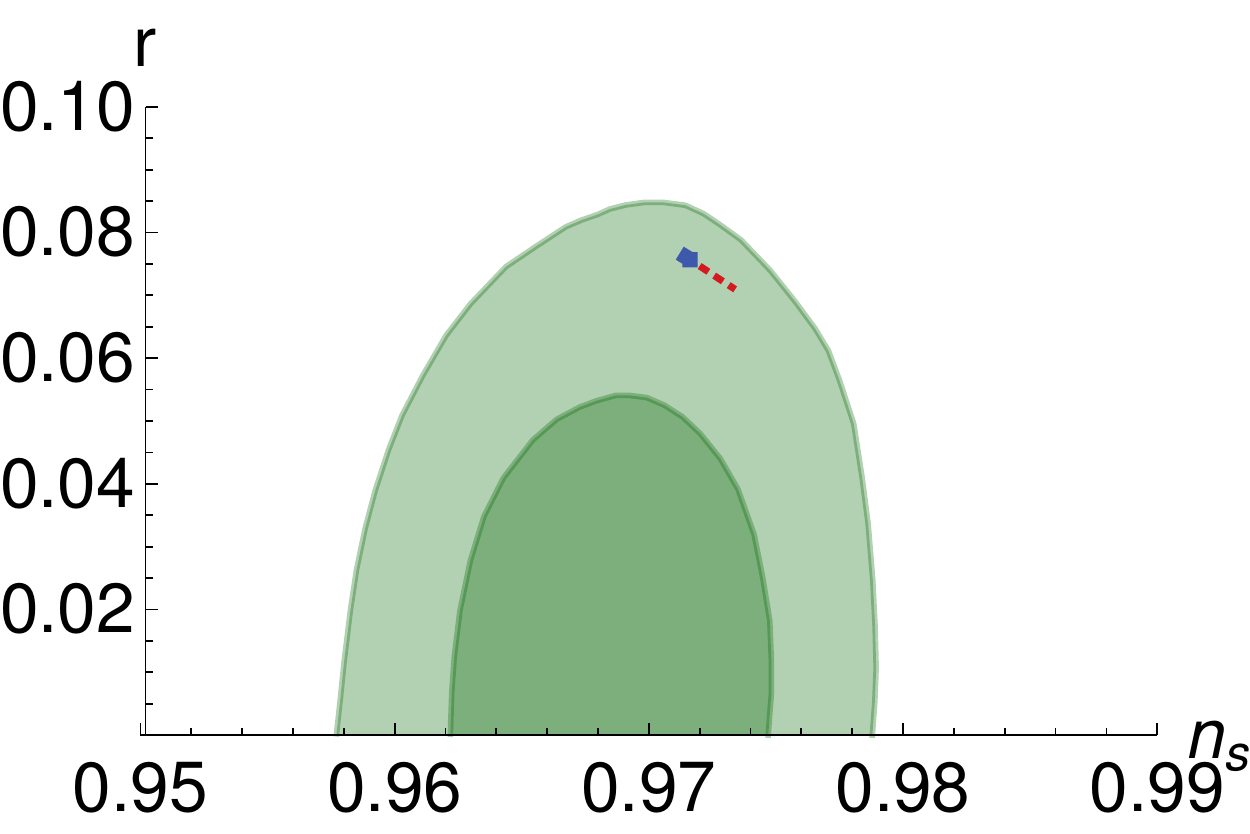}\\
  c) \hspace{0.45\textwidth} d)
 \end{center}
 \caption{Predictions in the metric (blue) and the Palatini (red, dashed) for number of $e$-folds $N_e$ as a function of $\xi_\phi$ (a), $\xi_\phi$ vs. $n_s$ (b), $r$  vs. $\xi_\phi$ (c) and  $r$ vs. $n_s$ (d). The light green areas present the 1,2 $\sigma$ constraints from the BICEP2/Keck data \cite{Array:2015xqh}. This figure is adapted with permission from \cite{Array:2015xqh}.}
 \label{Fig:Results:after}
\end{figure}

Now that we have determined the inflaton decay width, we can compute the effective number of $e$-folds, $N_e$. Since the inflaton decay widths in the metric and in the Palatini formulation differ (because of the different $q$ values), we expect different reheating temperatures and therefore different $N_e$.
The number of $e$-folds can be very well approximated as \cite{Ellis:2015pla} 
\begin{equation}
N_e \simeq 61.1181 + \frac{\left(1-3 \omega _{\text{int}}\right) \log \left(\frac{\rho _{\text{RH}}}{\rho
   _{\text{end}}}\right)}{12 \left(\omega _{\text{int}}+1\right)}+\frac{1}{4} \log
   \left(\frac{V_*^2}{\rho _{\text{end}} M_P^4}\right) \, , \label{eq:Ne}
\end{equation}
with
\begin{eqnarray}
 \omega_\text{int} &\simeq& \frac{0.782}{\log \left(\frac{2.096 m_{\phi }}{\Gamma _{\phi }}\right)} \, , \\
 \rho_{RH} &\simeq& 0.0151 \, \Gamma _{\phi }^2 M_P^2 \, ,
\end{eqnarray}
where $\omega_\text{int}$ and $\rho_{RH}$ are respectively the $e$-fold average of the equation-of-state parameter during the thermalization epoch and the energy density after reheating.
We focused on the linear limit to express the inflationary energy density $V_*$ and the energy density at the end of inflation $\rho_\text{end}$ in function of $N_e$ \cite{Ellis:2015pla}:
\begin{eqnarray}
 V_* &=& \frac{24 \pi ^2 A_s M_P^4}{4 N_e+1} \, , \\
 \rho_\text{end} &\simeq& \frac{18 \pi ^2 A_s M_P^4}{\left(4 N_e+1\right){}^{3/2}} \, .
\end{eqnarray}
Once we assign a value to the non-minimal coupling $\xi_\phi$, we can solve numerically eq. (\ref{eq:Ne}). The results are shown in Fig. \ref{Fig:Results:after}a, where we plot $N_e$ in function of $\xi_\phi$ for the metric (blue) and the Palatini formulation (red, dashed). In the metric case, for $\xi_\phi$ increasing, $N_e$ saturates to a constant value because $q$ is saturating to $\sqrt\frac{1}{6}$. Instead, in the Palatini formulation, $N_e$ is increasingly arbitrarily and therefore we chose $10^4$ as a maximum reference value for $\xi_\phi$. Then, in Figs. \ref{Fig:Results:after}b,c,d we give respectively the results for $\xi_\phi$ vs. $n_s$, $r$ vs. $\xi_\phi$ and $r$ vs. $n_s$ based on the results shown in Figs. \ref{Fig:Results:before}a and \ref{Fig:Results:after}a. We can see, in particular from Figs. \ref{Fig:Results:after}b  and \ref{Fig:Results:after}c, that for $\xi_\phi \gtrsim 1$ it is possible to discriminate between the two different formulations of gravity and that the Palatini formulation predicts a smaller (larger) value for $r$ ($n_s$) than the metric one. Moreover, for $\xi_\phi \simeq 1$, the metric formulation predicts a tensor-to-scalar ratio $r_m \simeq 0.076$, while the Palatini case gives  $r_P \simeq 0.075$. The difference between those two values, $\Delta r\simeq 10^{-3}$, is larger than the expected future sensitivity of the CORE mission \cite{Remazeilles:2017szm}.

\section{Summary and conclusions} \label{sec:Summary}
The most recent BICEP2/Keck collaboration results~\cite{Array:2015xqh}, which finds linear inflation on the edge of the allowed region, motivated us to review our previous results on non-minimal CW inflation~\cite{Kannike:2015kda} in light of different formulations of gravity (metric or Palatini).

First we considered a model independent approach and we noticed that both formulations produce linear inflation as an attractor limit, with the only difference lying in a normalization factor. Therefore, without knowing any detail about reheating, the two formulations are in practise not distinguishable. This result is not general but is a consequence of the particular choice of the non-minimal coupling in eq. (\ref{eq:Jordan:Lagrangian}). 

Therefore we considered the most minimal reheating scenario where the inflaton decays via the non-minimal coupling to gravity into SM particles. We showed  that, in the minimal configuration, the reheating happens via the non-minimal coupling $\xi_\phi$ into pairs of electroweak bosons (Higgses, $Z$'s  and $W$'s). 

Then, using the results of \cite{Ellis:2015pla}, we computed the number of $e$-folds corresponding to each formulation and subsequently the corresponding predictions for the tensor-to-scalar ratio $r$ and the scalar spectral index $n_s$. We noticed that it is possible to clearly discriminate between the two formulations for $\xi_\phi \gtrsim 1$.

In case linear inflation is confirmed by data, the future experimental sensitivity \cite{Remazeilles:2017szm} should be able to discriminate between one formulation or the other.

\acknowledgments

The author thanks Martti Raidal, Luca Marzola and Syksy R{\"a}s{\"a}nen for useful discussions.
This work was supported by the Estonian Research Council grants IUT23-6, PUT1026 and by the ERDF Centre of Excellence project TK133.

\appendix 
\section{Effective potential and minimization} \label{appendix:potential}
In this Appendix we discuss the effective potential and the possible scenarios that are arising from its minimization. Given the scalar potential in eq. (\ref{eq:Veff:J}), the general equation for minimising it at $v_\phi \neq 0$ is
\begin{equation}
 \frac{1}{4} \beta_{\lambda_\phi} (v_\phi) + \lambda_\phi (v_\phi) =0 \, ,
\end{equation}
where $\beta_{\lambda_\phi} (\mu)= \mu \frac{\partial}{\partial \mu}\lambda_\phi (\mu)$ is the beta-function of the quartic coupling $\lambda_\phi (\mu)$.
Therefore, several possibilities are open according to how we solve the equation:
\begin{eqnarray}
\text{a)} & \beta_{\lambda_\phi} (v_\phi)=\lambda_\phi (v_\phi)=0 , \label{eq:RGE:MPCP}\quad\\
\text{b)} & \beta_{\lambda_\phi} (v_\phi) > 0, \ \lambda_\phi (v_\phi)<0 ,\label{eq:RGE:bound:cond}\\
\text{c)} & \beta_{\lambda_\phi} (v_\phi) < 0, \ \lambda_\phi (v_\phi)>0  .
\end{eqnarray}
Further studies show that c) is actually a local maximum of the potential, therefore the only allowed possibilities are a) or b). Using eq. (\ref{eq:Vmin}), the first option implies also that $\Lambda=0$, realizing a full classical scale invariant setup, while the second option requires $\Lambda \neq 0$. 
The quartic coupling pre-factor in eq. (\ref{eq:Veff:J}) can be model-independently written as a Taylor series
\begin{equation}
\lambda_\phi(\phi) = \lambda_\phi(v_\phi) + 
                                  \beta_{\lambda _{\phi}}(v_\phi) \ln\frac{\phi }{v_\phi}+
                     \frac{1}{2!} \beta_{\lambda _{\phi}}'(v_\phi) \ln^2\frac{\phi }{v_\phi} +
                     \frac{1}{3!} \beta _{\lambda _{\phi}}''(v_\phi)  \ln^3\frac{\phi }{v_\phi} + \cdots ,
  \label{eq:lambdaTaylor}
\end{equation}
where $\beta_{\lambda _{\phi}}'(\mu)$ and $\beta _{\lambda _{\phi}}''(\mu)$ are respectively the first and second derivative of $\beta_{\lambda _{\phi}}(\mu)$ with respect to $t=\ln\mu$. Therefore for case a) described in eq. (\ref{eq:RGE:MPCP}) we have that the leading order expression is 
\begin{equation}
\lambda_\phi^a(\phi) \simeq  \alpha \ln^2\frac{\phi }{v_\phi} \, , \label{eq:lambda:run:2}
\end{equation}
where $\alpha=\frac{1}{2} \beta_{\lambda _{\phi}}'(v_\phi)$, while for case b) we get
\begin{equation}
\lambda_\phi^b(\phi) \simeq \lambda_\phi(v_\phi) + \beta_{\lambda _{\phi}}(v_\phi) \ln\frac{\phi }{v_\phi} \, ,
\end{equation}
which is the same equation as eq. (\ref{eq:lambda:run}). Since case b) has been extensively studied in the main body of the article, from now on we focus only on case a). Using eqs. (\ref{eq:Veff:J}) and (\ref{eq:lambda:run:2}) we get that the scalar potential in the Jordan frame is 
\begin{equation}
V_{\rm eff}^a = \alpha \ln^2 \frac{\phi}{v_\phi} \phi^4 \, . 
\end{equation}
Moving to the Einstein frame we have
\begin{equation}
V_{\rm eff, E}^a = \alpha \frac{M_P^4}{\xi_\phi^2}\ln^2 \frac{\phi}{v_\phi}  = \alpha M_P^2 \frac{q^2}{\xi_\phi^2} \phi_E^2  \, ,
\end{equation}
where we used eq. (\ref{eq:gE}) and eq. (\ref{eq:dphiE}).
Therefore, in the Einstein frame the inflaton potential is quadratic for both the formulations. 
The only difference is the normalization factor $q$, i.e., the value of the non-minimal coupling that satisfies the constraint (\ref{eq:As}) on the amplitude of scalar perturbations
\cite{Ade:2015xua,Ade:2015lrj}. We have seen in Section \ref{section:after:reheating} that is possible to distinguish between the two formulations by computing reheating and therefore predicting a different number of $e$-folds. However from \cite{Array:2015xqh} we see that quadratic inflation is ruled out at 2$\sigma$ for any number of $e$-folds, therefore this scenario is ruled out in both the formulations.

\section{Einstein frame potential and linear limit} \label{appendix:linear:limit}
\begin{figure}[t!]
\begin{center}
 \includegraphics[width=0.45\textwidth]{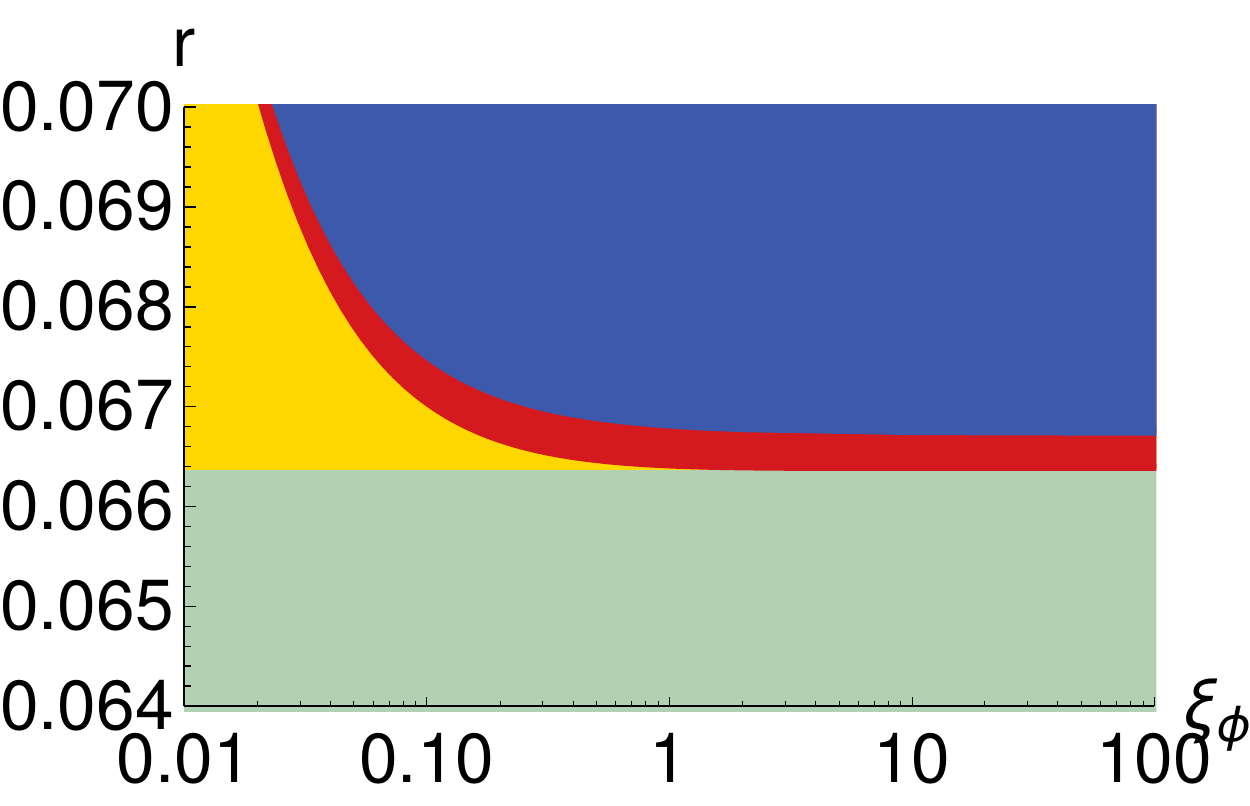}
 \end{center}
 \caption{Zoom in the $\xi_\phi \gg 1$ region of Fig. \ref{Fig:Results:before}b}
\label{Fig:r:vs:xi:zoom}
\end{figure}

As we noticed that in Figs.~\ref{Fig:Results:before}a and \ref{Fig:Results:before}b it seems impossible to distinguish between the metric and the Palatini formulation. Both models are saturating the linear limit for $\xi_\phi\gtrsim 0.1$, however their limits slightly differ. This can only be appreciated by zooming in the $\xi_\phi \gg 1$ region in Fig.~\ref{Fig:Results:before}b. We do this in Fig. \ref{Fig:r:vs:xi:zoom}. As we can see both models are saturating to a fixed $r$ value, however such value is slightly different. The reason is the following. In metric formulation, for $\xi_\phi \to \infty$ we have $q \to \sqrt\frac{1}{6}$, therefore, the asymptotic potential would be 
\begin{equation}
V_E(\phi_E)=\Lambda ^4 \left(4 \, \sqrt\frac{1}{6} \frac{\phi_E}{M_P}+e^{-4 \, \sqrt\frac{1}{6} \frac{\phi_E}{M_P}}-1\right) \, ,
\end{equation}
which is, in the slow-roll region (i.e. $\phi_E \gg M_P$) very close (but different) to a linear potential. On the other side in the Palatini formulation, for $\xi_\phi \to \infty$ we obtain $q \to \infty$, and therefore the asymptotic potential is exactly a linear one: $V \approx \sqrt\xi_\phi \Lambda^4 \frac{\phi_E}{M_P}$.

\bibliographystyle{JHEP}
\bibliography{citations}

\end{document}